\documentclass{article}

\usepackage[preprint]{neurips_2019}

\usepackage[utf8]{inputenc} 
\usepackage[T1]{fontenc}    
\usepackage{hyperref}       
\usepackage{url}            
\usepackage{booktabs}       
\usepackage{amsfonts}       
\usepackage{nicefrac}       
\usepackage{microtype}      
\usepackage{mathrsfs,amsmath}
\usepackage{graphicx}

\usepackage{caption}
\usepackage{subcaption}

\usepackage[colorinlistoftodos]{todonotes}

\newcommand{\A}{\mathcal{A}}

\newcommand{\n}{\mathbf{n}}
\newcommand{\dd}{\mathbf{d}}
\newcommand{\p}{\mathbf{p}}
\newcommand{\s}{\mathbf{s}}

\newcommand{\m}{\mathbf{m}}

\LetLtxMacro{\originaleqref}{\eqref}
\renewcommand{\eqref}{Eq.~\originaleqref}

\title{i-RIM applied to the fastMRI challenge}
\author{Patrick Putzky \\
AMLAB, University of Amsterdam \\
MPI for Intelligent Systems, T\"ubingen \And Dimitrios Karkalousos \\
AMLAB, University of Amsterdam 
\And Jonas Teuwen \\
Radboud University Medical Center \\
Netherlands Cancer Institute
\And Nikita Miriakov \\
Radboud University Medical Center \And Bart Bakker \\Philips Research, The Netherlands \And Matthan Caan \\ Amsterdam UMC, University of Amsterdam \\ Dept. of Biomedical Engineering and Physics \And Max Welling \\ AMLAB, University of Amsterdam \\ Canadian Institute for Advanced Research}
\begin{document}
\maketitle
\vspace{-.4cm}
\begin{abstract}
    We, team AImsterdam, summarize our submission to the fastMRI challenge \citep{zbontar2018fastmri}. Our approach builds on recent advances in invertible ``learning to infer'' models as presented in \citet{putzky2019}. Both, our single-coil and our multi-coil model share the same basic architecture.
\end{abstract}

\section{Introduction}
To solve the accelerated MRI problem as presented in the fastMRI challenge \citep{zbontar2018fastmri}, we train an invertible Recurrent Inference Machine (i-RIM) for each of the challenges \citep{putzky2019}. The i-RIM is an invertible variant of the RIM \citep{putzky2017recurrent} which has been successfully applied to accelerated MRI before \citep{Lonning2019}. The formulation of the i-RIM allows us to stably train models which are several hundreds of layers deep. Most of our approach is already described in \citet{putzky2019}. Here, we will focus on changes to \citet{putzky2019} which were done for the challenge, and on the adaptation to the multi-coil setting.

We treat the problem of accelerated MRI as an inverse problem with a forward model given by
\begin{align}
\dd^{(i)} &= \mathbf{P}\mathscr{F} \p^{(i)} + \n^{(i)}
\end{align}
where $\dd^{(i)} \in \mathbb{C}^m$ are sub-sampled k-space measurements at coil $i$, $\mathbf{P}$ is a sampling mask, $\mathscr{F}$ is a Fourier transform, $\p^{(i)} \in \mathbb{C}^n$ is an image recorded at coil $i$, and $\n^{(i)}$ is the noise at coil $i$. In our approach, we do not explicitly model spatial coil sensitivity maps as is commonly done in other approaches. We stack k-space measurement and coil images from all coils, respectively, such that the forward model takes the form
\begin{align}\label{eq:mri_measurement}
\dd &=  \A \p + \n
\end{align}
with
\begin{equation}
  \begin{aligned}
  \dd = \begin{pmatrix}\dd^{(1)}\\ \vdots\\ \dd^{(K)}\end{pmatrix}
  \end{aligned}
  \qquad
  \begin{aligned}
  \p = \begin{pmatrix}\p^{(1)}\\ \vdots\\ \p^{(K)}\end{pmatrix}
  \end{aligned}
  \qquad
  \begin{aligned}
  \n = \begin{pmatrix}\n^{(1)}\\ \vdots\\ \n^{(K)}\end{pmatrix}
  \end{aligned}
  \qquad
  \begin{aligned}
  \A = \mathrm{1}_K \otimes \mathbf{P}\mathscr{F}
  \end{aligned}
\end{equation}
where $\otimes$ denotes the Kronecker product, $K$ is the total number of coils in the problem, i.e. $15$ in the multi-coil setting, and $1$ in the single-coil setting.
\section{Method}
The i-RIM is a deep learning model which iteratively updates its machine state $(\p_t,\s_t)$ based on simulations of the forward model in \eqref{eq:mri_measurement} such that
\begin{align} \label{eq:iterative_models}
\p_{t+1}, \s_{t+1} = h_\phi(\A,\dd,\p_t,\s_t)
\end{align}
where $\p_{t}$ is the models estimate of $\p$ and $\s_t$ is a latent state at iteration $t$, respectively. Many modern approaches to solving inverse problems which we refer to as ``learning to infer'' models can be summarized through equation \eqref{eq:iterative_models}. What differentiates the i-RIM from other approaches is that (1) the only model assumption is in the forward model which makes the i-RIM a mostly data-driven approach, and (2) $h_\phi$ is fully invertible which allows us to train the model with back-propagation without storing intermediate activations \citep{Gomez2017}. Hence, we can train arbitrarily deep networks. Empirical results in deep learning suggest that deeper models almost always perform better than their shallow counterparts \citep{He2015}. The i-RIM brings this potential to ``learning to infer'' models.

For the i-RIM, \eqref{eq:mri_measurement} specifically takes the form
\begin{align} \label{eq:rim_update}
\p_{t+1}, \s_{t+1} = h_\phi(\nabla\mathcal{D}\left(\dd, \A, \p_t\right),\p_t,\s_t)
\end{align}
where
\begin{align*}
    \nabla\mathcal{D}\left(\dd, \A, \p_t \right) = \A^H \left(\A \p_t - \dd \right)
\end{align*}
is the gradient of the data consistency term under a normal likelihood model with $\A^H$ being the adjoint operator of $\A$. This gradient reflects how well the current estimate is supported by the measured data under the forward model.
To produce the final estimate of $\p$ we use a non-invertible block such that
\begin{align}
    \hat{\p} = f_\theta(\p_T,\s_T)
\end{align}
is the models final complex-valued estimate with $\hat{\p} \in \mathbb{C}^n$. The competition results are evaluated on magnitude images, hence we do $\hat{\m} = |\hat{\p}|$ to generate magnitude images for the competition. As training loss we use the structural similarity loss \citep{Wang2004}:
\begin{align}
    \mathcal{L}(\phi, \theta) = -\frac{1}{N} \sum_{j=1}^N \operatorname{SSIM}(\hat{\m}_j,\m_j)
\end{align}
where $N$ is the mini-batch size. As the initial machine state we set
\begin{align}
\p_0 &= \A^H \dd \\
\s_0 &= \begin{pmatrix}\omega \\ \mathbf{0}_{D-8}\end{pmatrix}
\end{align}
where $\p_0$ is the zero-filled corrupted image, and $\omega$ is a 1-hot vector which encodes meta-data about the experimental condition such as field strength (1.5T vs. 3T) and fat suppressed vs. non-fat suppressed data. This meta-data can potentially activate different pathways in the i-RIM under the different experimental conditions.


\paragraph{Models}
We trained separate models for the single-coil and multi-coil challenges with 8 steps each. At each step, the models have 12 down-sampling blocks 
(see \citet{putzky2019}). In total, this amounts to 480 layer deep networks. The single-coil model has a machine state of 64 feature layers, and the multi-coil model has a machine state of 96 feature layers.

\paragraph{Training}
Because the volumes in the data set have vastly different sizes, we cropped the central portion of each image slice to $368 \times 368$ pixels. For smaller slices we applied zero padding to produce slices of the appropriate size. We simulated k-space measurements using the sampling mask function supplied by \citet{zbontar2018fastmri} with $4\times$ and $8\times$ acceleration factors. As target images we used ESC images for the single-coil model and RSS targets for the multi-coil model, respectively (see \cite{zbontar2018fastmri}). We used the Adam optimizer with initial learning rate $10^{-4}$ which was reduced by factor $10$ every 30 epochs.

\section{Results}
\begin{figure}
     \centering
     \begin{subfigure}[b]{0.3\textwidth}
         \centering
         \caption{Multi-Coil RSS \\\hspace{\textwidth}Ground Truth}
         \includegraphics[width=\textwidth]{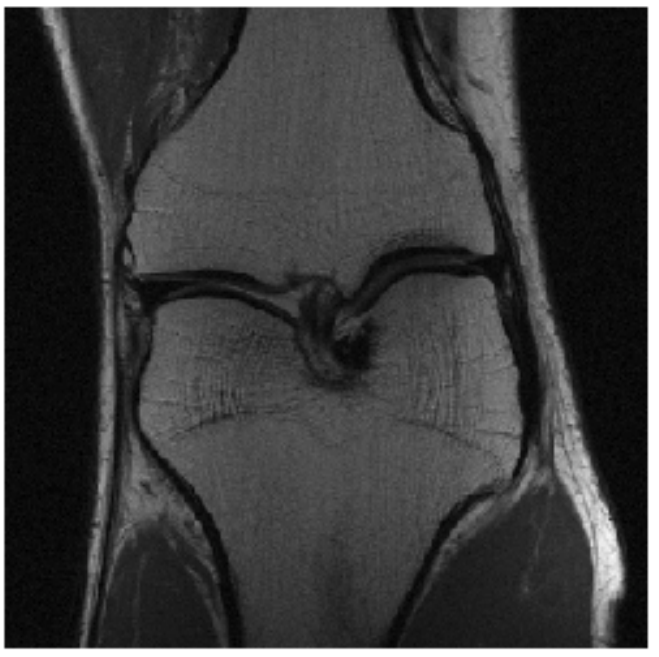}
         \label{fig:multicoil_val_file1000126_slice_18_reconstruction_rss}
     \end{subfigure}
     \hfill
     \begin{subfigure}[b]{0.3\textwidth}
         \centering
         \caption{Multi-Coil 4-Fold \\\hspace{\textwidth}Reconstruction}
         \includegraphics[width=\textwidth]{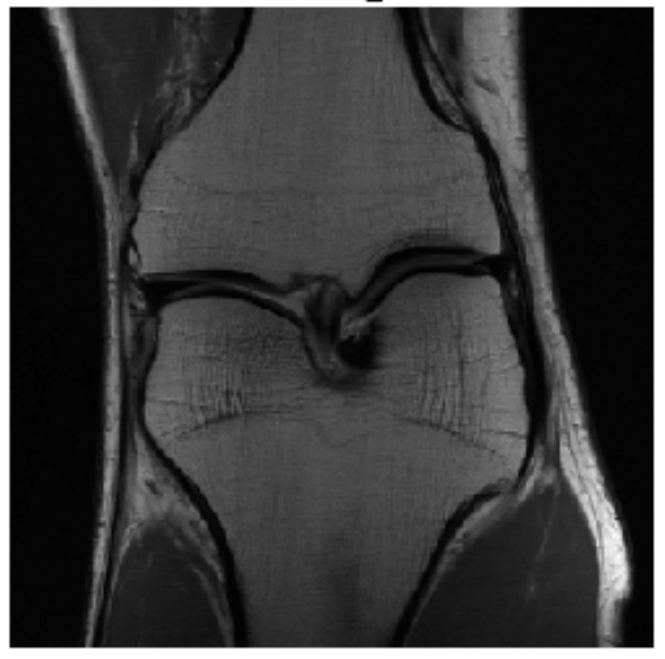}
         \label{fig:mutlicoil_val_file1000126_slice_18_irim_reconstruction_4x}
     \end{subfigure}
     \hfill
     \begin{subfigure}[b]{0.3\textwidth}
         \centering
         \caption{Multi-Coil 8-Fold \\\hspace{\textwidth}Reconstruction}
         \includegraphics[width=\textwidth]{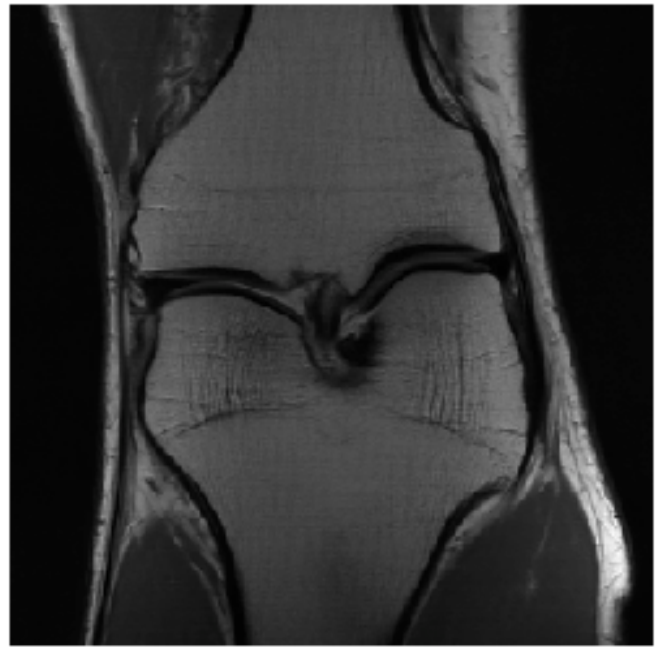}
         \label{fig:mutlicoil_val_file1000126_slice_18_irim_reconstruction_8x}
     \end{subfigure}
     \\
     \begin{subfigure}[b]{0.3\textwidth}
         \centering
         \caption{Single-Coil ESC \\\hspace{\textwidth}Ground Truth}
         \includegraphics[width=\textwidth]{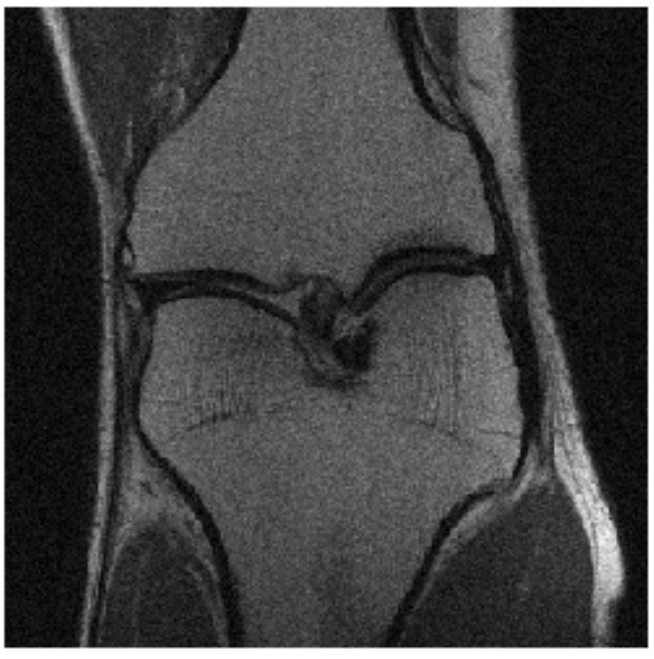}
         \label{fig:singlecoil_val_file1000126_slice_18_reconstruction_esc}
     \end{subfigure}
     \hfill
     \begin{subfigure}[b]{0.3\textwidth}
         \centering
         \caption{Single-Coil 4-Fold \\\hspace{\textwidth}Reconstruction}
         \includegraphics[width=\textwidth]{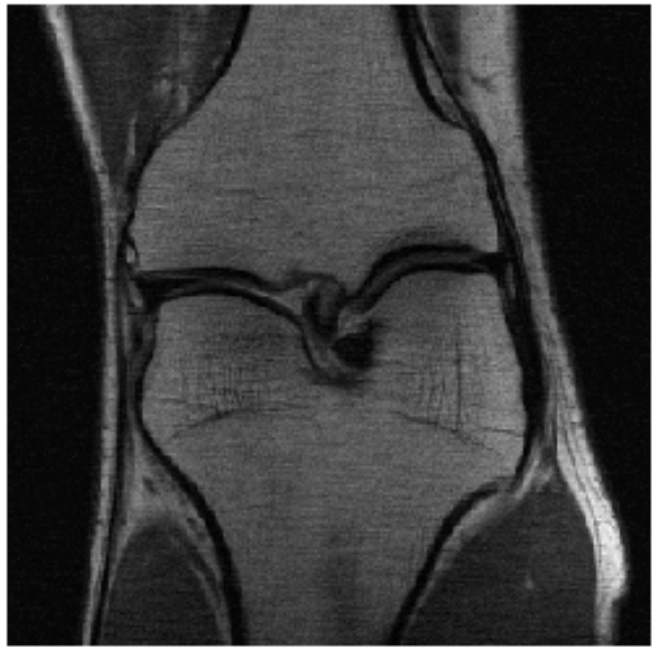}
         \label{fig:singlecoil_val_file1000126_slice_18_irim_reconstruction_4x}
     \end{subfigure}
     \hfill
     \begin{subfigure}[b]{0.3\textwidth}
         \centering
         \caption{Single-Coil 8-Fold \\\hspace{\textwidth}Reconstruction}
         \includegraphics[width=\textwidth]{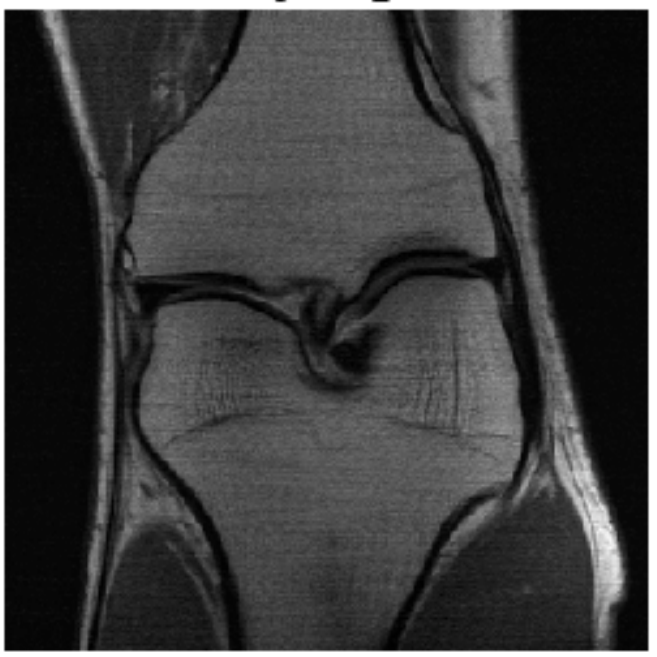}
         \label{fig:singlecoil_val_file1000126_slice_18_irim_reconstruction_8x}
     \end{subfigure}     
    \caption{Example reconstructions. The reconstructions visually improve the ground truth images, suggesting a strong prior.}
    \label{fig:reconstructions}
\end{figure}

We evaluated our models on three data sets: the validation set as in \citet{zbontar2018fastmri}, and the test and challenge sets through the fastMRI website. A summary of these evaluations can be found in table \ref{tab:fastmri}\footnote{Results on the challenge data set will be added once publicly available.}. To assess image quality more closely, we show some exemplary reconstructions from each model in figure \ref{fig:reconstructions}.

\begin{table}[t]
\centering
\caption{Reconstruction performance on different data sets from the fastMRI challenge \cite{zbontar2018fastmri} under different metrics. NMSE - normalized mean-squared-error; PSNR - peak signal-to-noise ratio; SSIM - structural similarity index \cite{Wang2004}. $\downarrow$ - lower is better; $\uparrow$ higher is better.}
\begin{tabular}{@{}lcccccc@{}}
\toprule
& \multicolumn{3}{c}{4x Acceleration}  & \multicolumn{3}{c}{8x Acceleration}           \\
\cmidrule(r){2-4} \cmidrule(r){5-7}
\textbf{i-RIM single-coil} & NMSE $\downarrow$& PSNR $\uparrow$& SSIM $\uparrow$& NMSE $\downarrow$& PSNR $\uparrow$& SSIM $\uparrow$\\
\cmidrule(r){1-1} \cmidrule(r){2-4} \cmidrule(r){5-7}
Validation & $0.0342$& $32.43$& $0.751$& $0.0446$ & $30.92$ & $0.692$\\
Test & $0.0272$ & $33.65$ & $0.781$ & $0.0421$ & $30.56$ & $0.687$ \\
Challenge & & & & & & \\
\midrule
\textbf{i-RIM multi-coil} & NMSE $\downarrow$& PSNR $\uparrow$& SSIM $\uparrow$& NMSE $\downarrow$& PSNR $\uparrow$& SSIM $\uparrow$\\
\cmidrule(r){1-1} \cmidrule(r){2-4} \cmidrule(r){5-7}
Validation & $0.0062$& $38.84$ & $0.916$ & $0.0103$ & $36.19$ & $0.886$ \\
Test & $0.0052$ & $39.52$ & $0.928$ & $0.0093$& $36.53$ & $0.887$ \\
Challenge & & & & & & \\
\bottomrule
\end{tabular}
\label{tab:fastmri}
\end{table}

\subsubsection*{Acknowledgements}
Patrick Putzky and Dimitrios Karkalousos were supported by Philips Research.

\bibliographystyle{unsrtnat}
\bibliography{references}

\end{document}